\documentclass[review]{elsarticle}

\usepackage{lineno,hyperref}
\modulolinenumbers[5]

\journal{Phisycs Letters A}

\begin{document}

\begin{frontmatter}

\title{Neutron emission from fracturing of granite blocks: An experimental revisitation}
%\title{Elsevier \LaTeX\ template\tnoteref{mytitlenote}}



\author[chimica-address]{P.~Benetti}
\author[infn-address]{F.~Boffelli}
\author[infn-address]{C.~Marciano\fnref{footnote}}
\author[fisica-address]{A.~Piazzoli}
\author[infn-address]{G.L.~Raselli\corref{mycorrespondingauthor}}
\cortext[mycorrespondingauthor]{Corresponding author}
\ead{gianluca.raselli@pv.infn.it}

\fntext[myfootnote]{Former address of C. Marciano.}


\address[chimica-address]{Dipartimento di Chimica, Universit\`a di Pavia,  Via Torquato Taramelli, 12, Pavia, Italy}
\address[infn-address]{INFN Sezione di Pavia, Via A. Bassi 6, Pavia, Italy}
%\address[marciano-address]{Private consultant, formerly  INFN Sezione di Pavia, Via A. Bassi 6, Pavia, Italy}
\address[fisica-address]{Dipartimento di Fisica, Universit\`a di Pavia, Via A. Bassi 6, Pavia, Italy}

\begin{abstract}
A series of experimental tests, such as those of Carpinteri et al. (2013), have been performed. The aim was to check the emission of neutrons in the fracture of Luserna granite blocks under mechanical loading, as reported by the above mentioned authors. No neutrons have been detected and some doubts have emerged on the soundness of the previous measurements.

\end{abstract}

\begin{keyword}
{Neutron emissions \sep Neutron detection \sep Piezonuclear reactions \sep Rocks crushing failure}
%\texttt{elsarticle.cls}\sep \LaTeX\sep Elsevier \sep template
%\MSC[2010] 00-01\sep  99-00
\end{keyword}

\end{frontmatter}

%\linenumbers

\section{Introduction}
\label{intro}

A paper on ''Piezonuclear Fission Reactions'' was published in the year 2013 \cite{Carpinteri}. In this work the authors stated that neutrons are emitted from brittle rock specimen under mechanical loading while crashing. The obtained experimental data were 
thoroughly described and discussed along with some conclusions also involving geochemistry among other topics.
That work was the last one of a long series of papers, both experimental and theoretical, on piezonuclear fission reactions, published by the same authors or in collaboration with other groups investigating this subject (see  \cite{Carpinteri,Cardone} and references therein),
and can be considered as a summary of all the previous work done.
In this context it should also be underlined that several doubts have been reported
in literature on the performed neutron analysis and interpretation of the results~\cite{Spallone,Amato}.
More recently several of the papers published on this matter by Carpinteri et al. have been withdrawn \cite{Link}, but the one we are dealing with here was not.

Since the topic could potentially be of no-minor scientific relevance, %significant scientific relevance,
we decided to repeat the mechanical loading experiments, striving to use as much as possible the same technique, kind of sample and neutron detector employed in \cite{Carpinteri}. 

\begin{figure}[p]
\begin{center}
 \includegraphics[width=\textwidth]{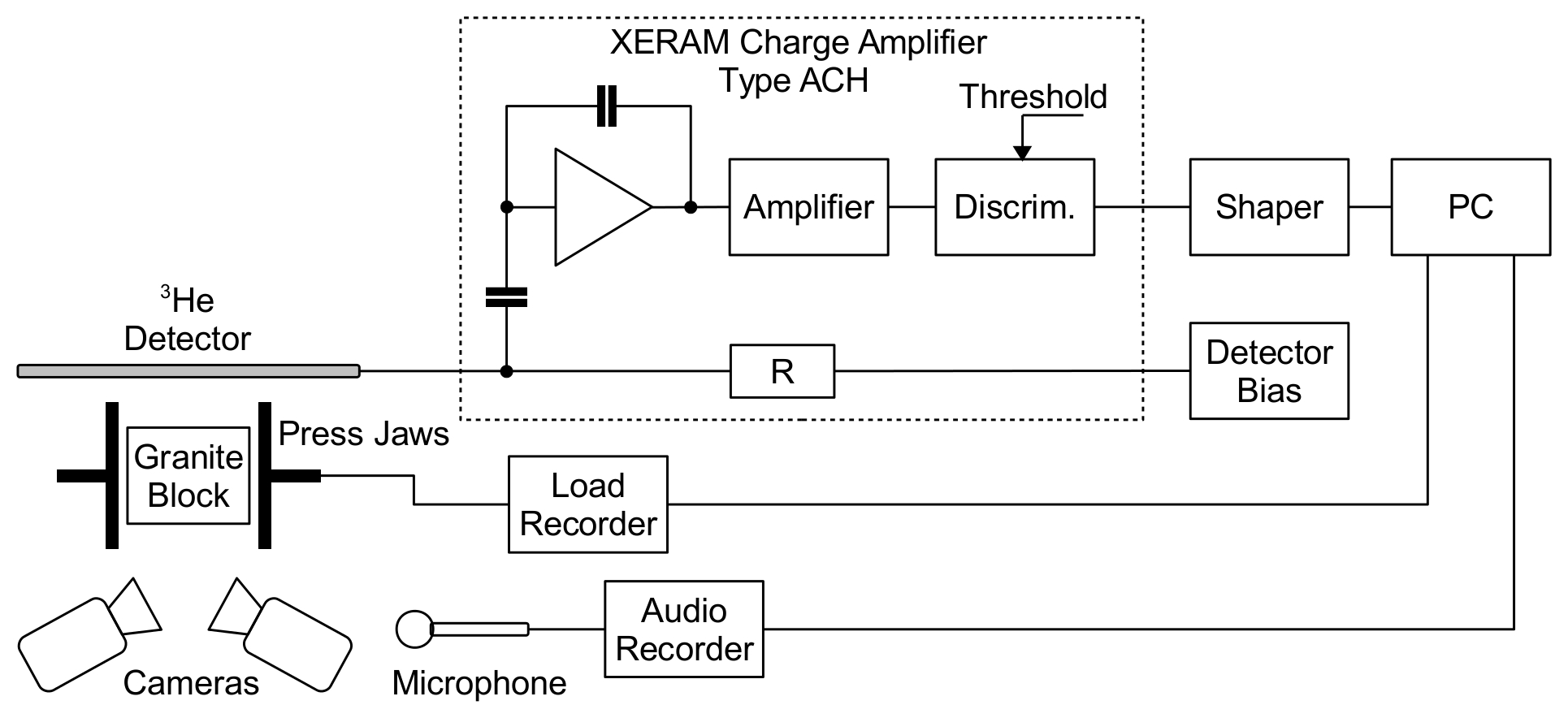}
\caption{Schematics of the experimental set-up used in the compression tests.}
\label{setupp}
\end{center}
\end{figure}

Our experimental set-up is sketched in Fig. \ref{setupp}.
The cubic specimens were Luserna granite, the same material used by Carpinteri et al. in their tests, having dimensions either
$10 \times 10 \times 10$~cm$^3$ or $10 \times 10 \times 8$~cm$^3$. No differences were observed between them, except the collapsing load. 
The neutron detector used in the compression tests is a $^3$He type 65-NH-45 manufactured by Xeram, (France) with its standard electronics \cite{Xeram1,Xeram2}. The detector (25~mm diameter, 450~mm long) is exactly of the same model of that used in \cite{Carpinteri}. Its thermal neutron sensitivity is 65 cps for a thermal neutron flux of 1 neutron s$^{-1}$~cm$^{-2}$. During the tests the detector was powered at 1.4~kV by means of a high voltage CAEN N126 power supply. The electronics performs an integral-differentiation of the $^3$He signal (1 $\mu$s integration, 4 $\mu$s differentiation) followed by a discrimination which supplies output TTL pulses (10 $\mu$s width). 
The time width of these pulses was stretched to 1~ms by means of a Gate Generator (CAEN 2255B), the output of which was directly connected to the input of a PC sound card board used as data-acquisition card allowing the sampling of the input pulses at 10~kSa/s and
%The adopted data acquisition software allowed the sampling of the input pulses at 10~kHz rate, 
%giving the shape and 
giving the time of occurrence of each $^3$He count with a precision of 0.1 ms.  The sensitivity of the $^3$He detector was checked before the measurements against a known neutron flux~\cite{Alloni}.
The press employed is a servo-hydraulic ``200 Mp Compression Testing Machine model 200 D 76'', manufactured by Amsler (Switzerland) \cite{Amsler}. The maximum value of the applied loading force is 2000 kN.
The adopted instrumentation was integrated with two video cameras and an audio recorder. The acquisition of $^3$He pulses, mechanical load,
video and audio signals was synchronized at the beginning of each run.

The minimal possible distance between the center of the rock sample and the detector was 26~cm. In comparison with \cite{Carpinteri}, there is a reduction of a factor six in the collected solid angle, nevertheless, according to the data in Table 2 of the same reference, more than one neutron should still be detected in the crash. 

\section{Results}
\label{Results}

In all we carried out six runs, each one consisting of a handful of tests.

\begin{figure}[p]
\begin{center}
\includegraphics[width=0.85\textwidth]{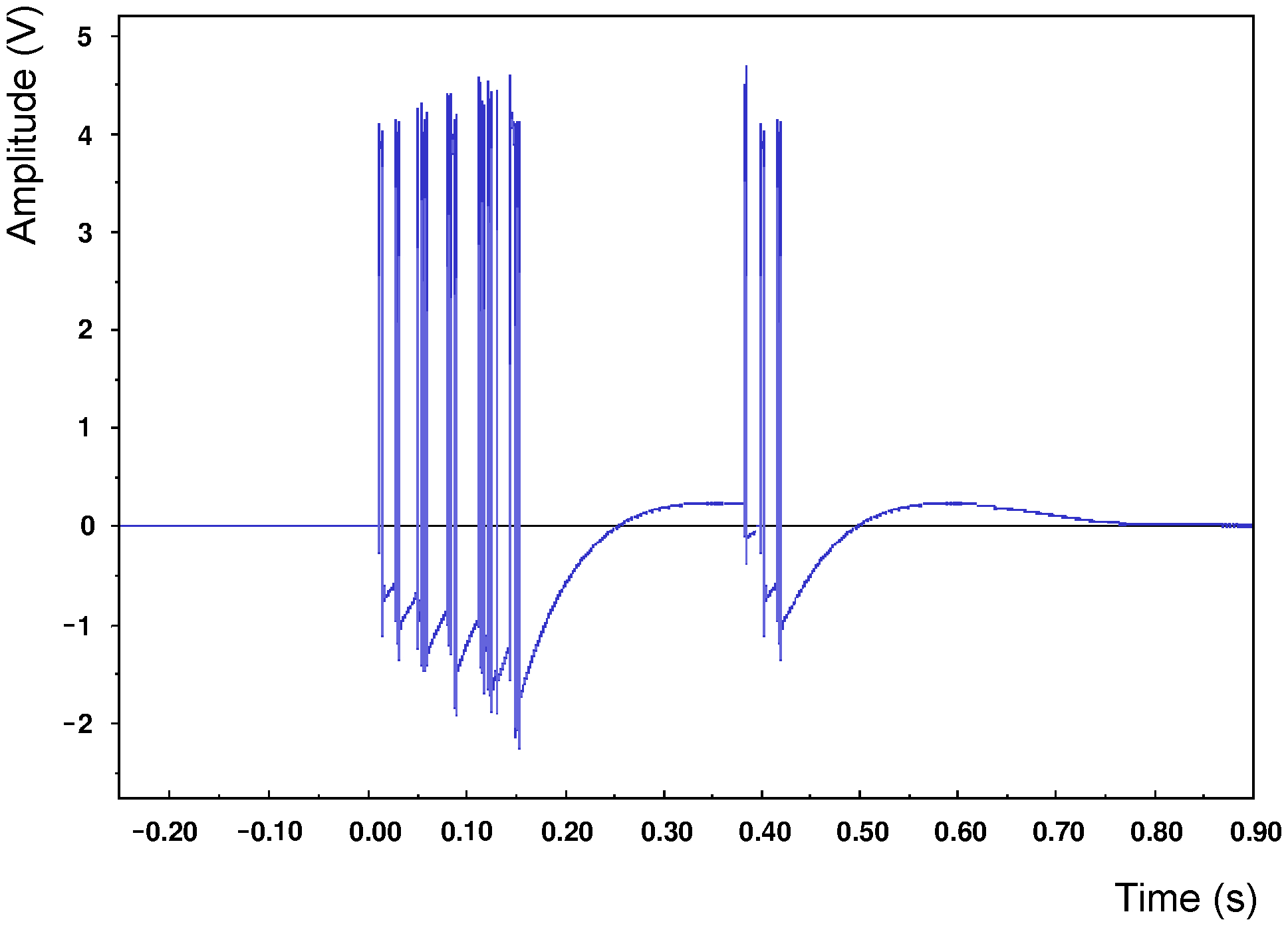}
\caption{Acquired $^3$He signals showing many TTL pulses surfing over a strongly perturbed baseline.}
\label{Fig2}  
\end{center}
\end{figure}

\begin{figure}[p]
\begin{center}
\includegraphics[width=0.85\textwidth]{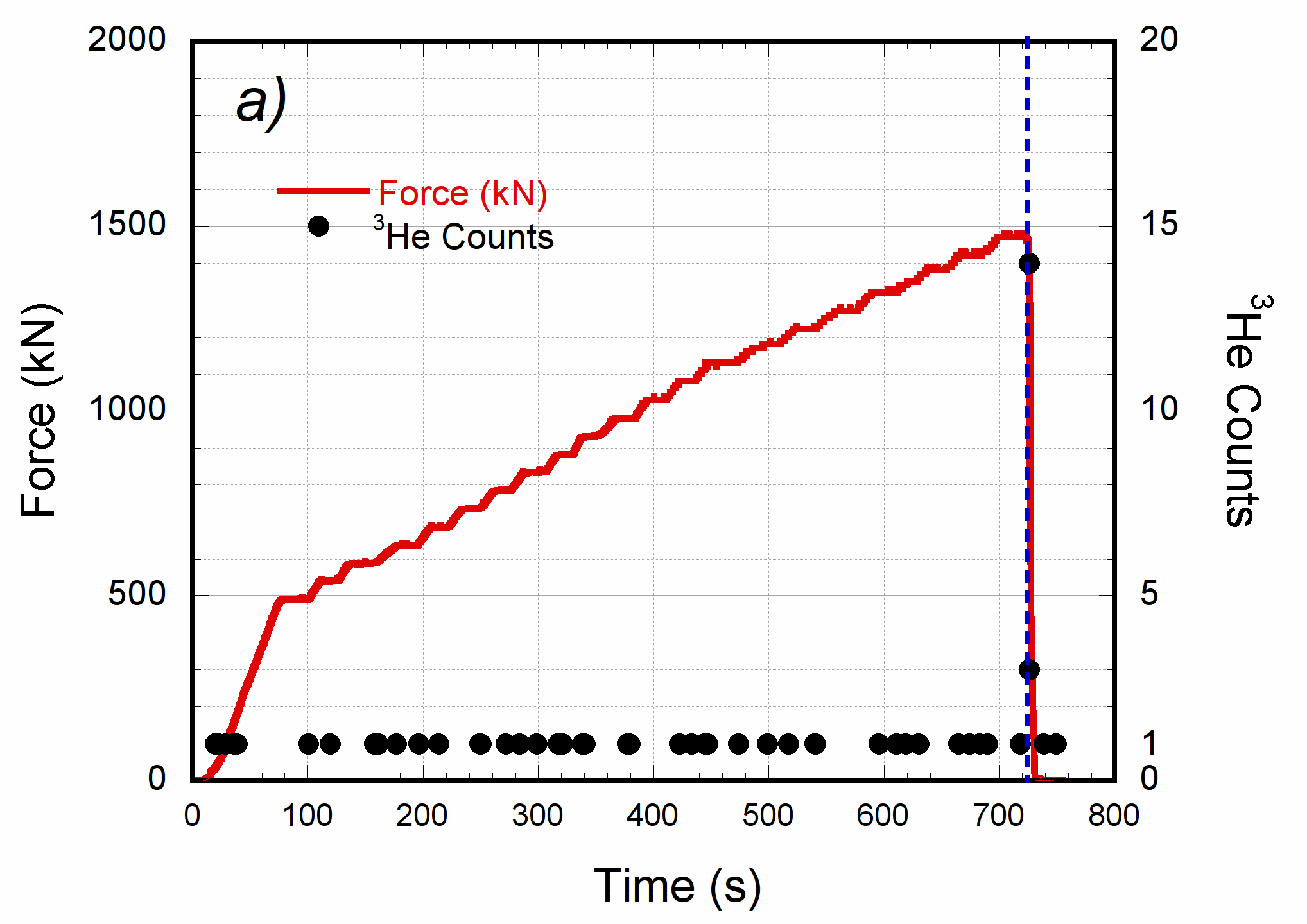}\\
\includegraphics[width=0.85\textwidth]{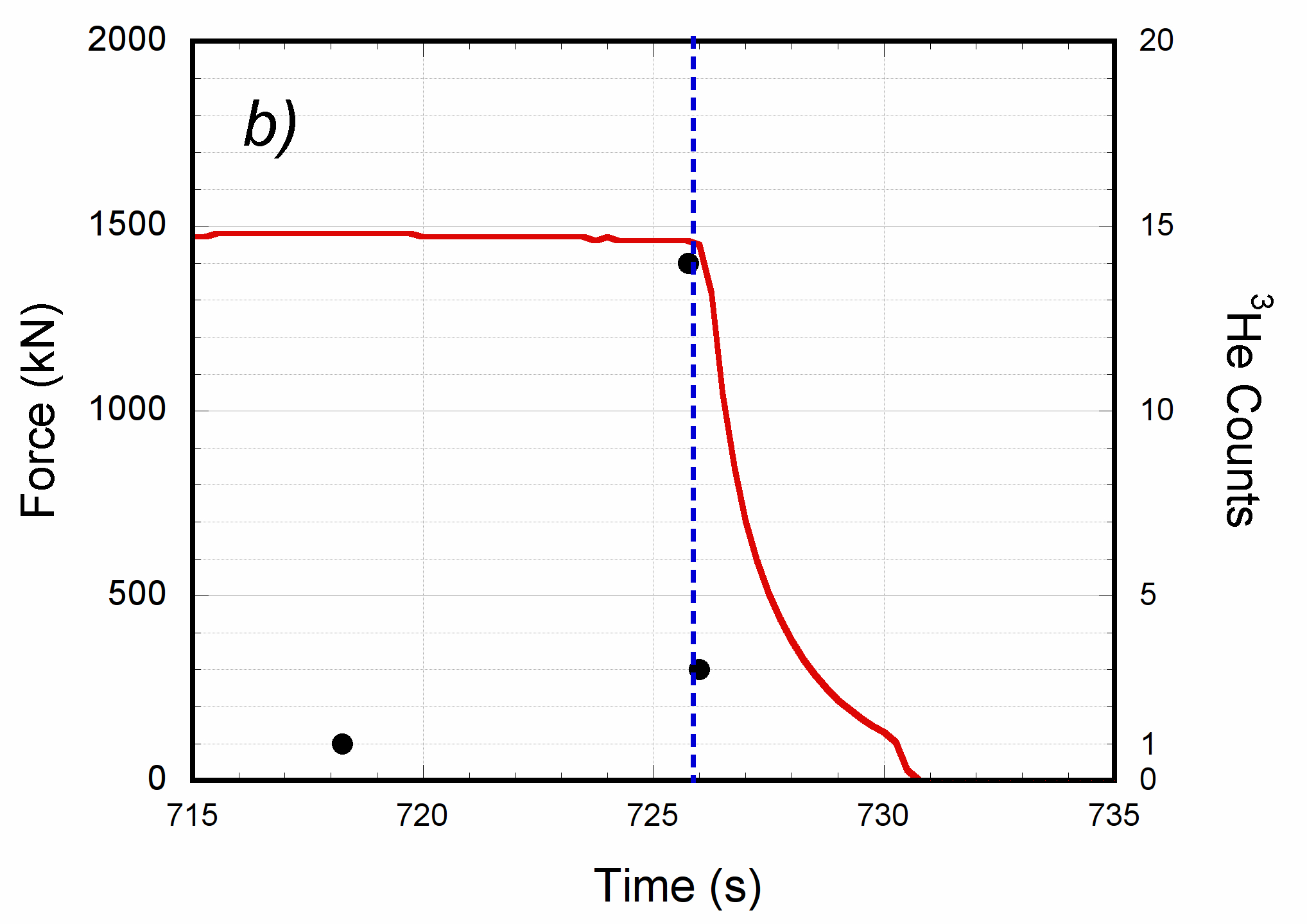}
\caption{$^3$He counts (black dots) together with the loading force on the specimen (red line) in one early test with the unshielded neutron detector. In a) the whole experimental test time is plotted whereas in b) a zoom around the crashing moment is represented. The vertical blue dashed line, in the plots, indicates the crashing time.}
\label{Fig3}
\end{center}
\end{figure}

In one of the earliest test, the $^3$He detector was accidentally hit by a mechanical component of the hydraulic press. The acquired signal, shown in Fig. \ref{Fig2}, presented a complex waveform in which many TTL signals were surfing over a strongly perturbed baseline. 
The corresponding $^3$He counts are shown as black dots in Fig.~\ref{Fig3}, together with the loading force exerted by the press on the specimen (red line). The vertical blue dashed line in the plots indicates the crashing time. At first glance these signals should have been associated to neutrons detection but, because of the mentioned shock accident and the high number of associated pulses,
%their long lasting, 
it was impossible to rule out a mechanical effect on electronics. Furthermore, even an acoustic effect was eligible, as supposed in~\cite{Spallone}.

\begin{figure}[p]
\begin{center}
\includegraphics[width=0.85\textwidth]{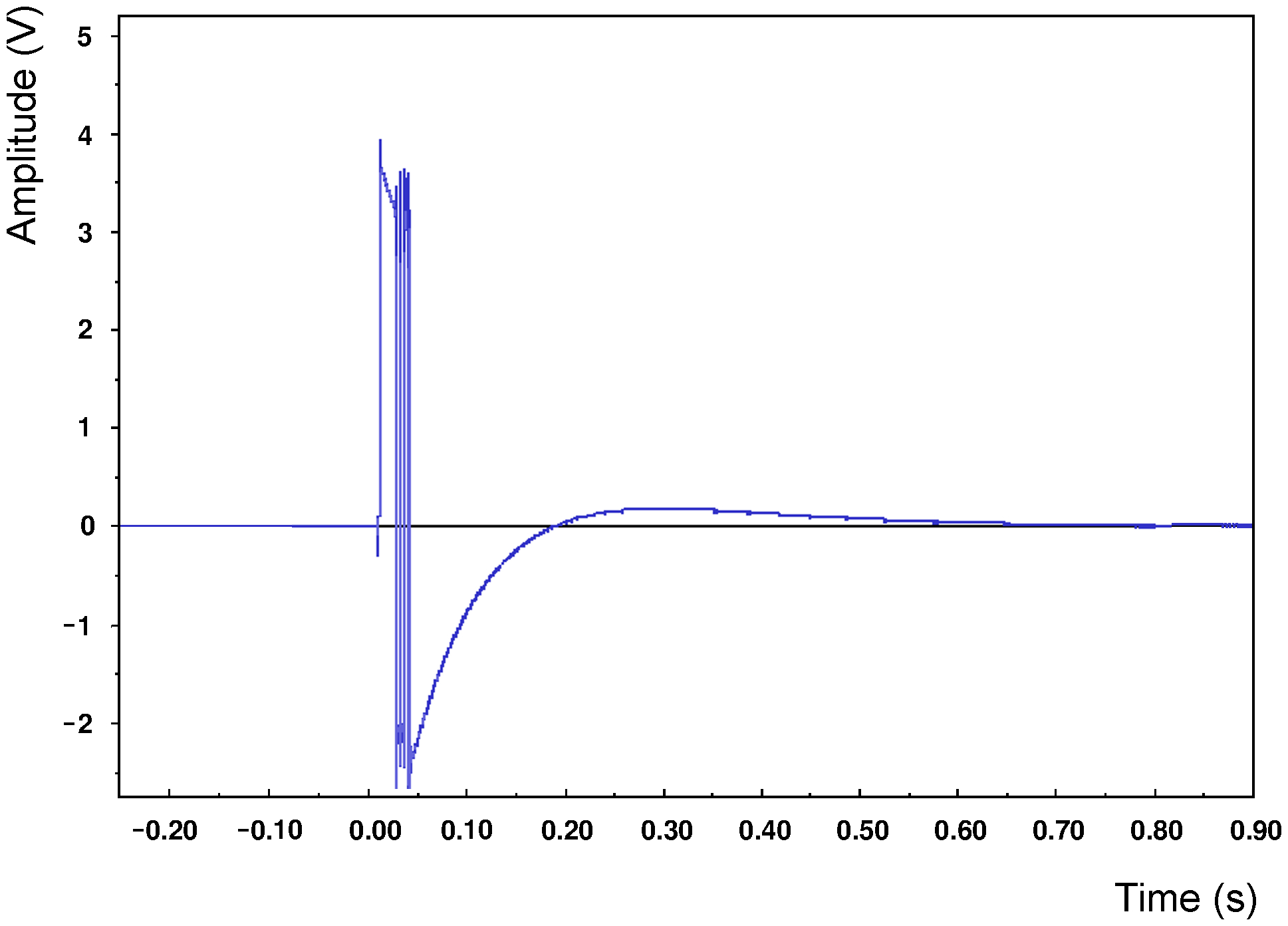}
\caption{TTL signals acquired when a blank gun was fired near the $^3$He detector.}
\label{Fig4}  
\end{center}
\end{figure}

This last hypothesis found a proof when we intentionally fired a blank gun near the $^3$He detector: TTL signals were recorded, exactly as in the case of neutron occurrence, as shown in Fig. \ref{Fig4}. %Since then we started a systematic investigation on this artifact in the detection unit, using the blank gun mentioned above and moving from a naked detector until a satisfactory acoustical shielding.
Following this finding, we started a systematic investigation of this artifact in the detection unit. To this purpose, we repeated the test with the blank gun mentioned above by installing an increasing amount of acoustical shielding material, until we found no additional pulses above the background single ones at all.

In this investigation it was a surprise to find out that the preamplifier connected to the $^3$He tube was very sensitive to the acoustic perturbation, even more than the tube itself. 

\begin{figure}[p]
\begin{center}
\includegraphics[width=0.85\textwidth]{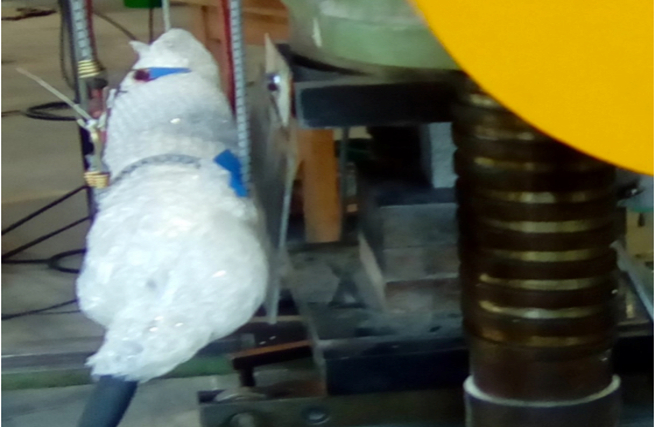}
\caption{The mechanical press, the granite block inside its jaws and the detector, suspended and wrapped in a white protective plastics. The $^3$He detection system was recalibrated against thermal neutrons with this protection in place.}
\label{Fig5}      
\end{center} 
\end{figure}

\begin{figure}[p]
\begin{center}%[h]
\includegraphics[width=0.85\textwidth]{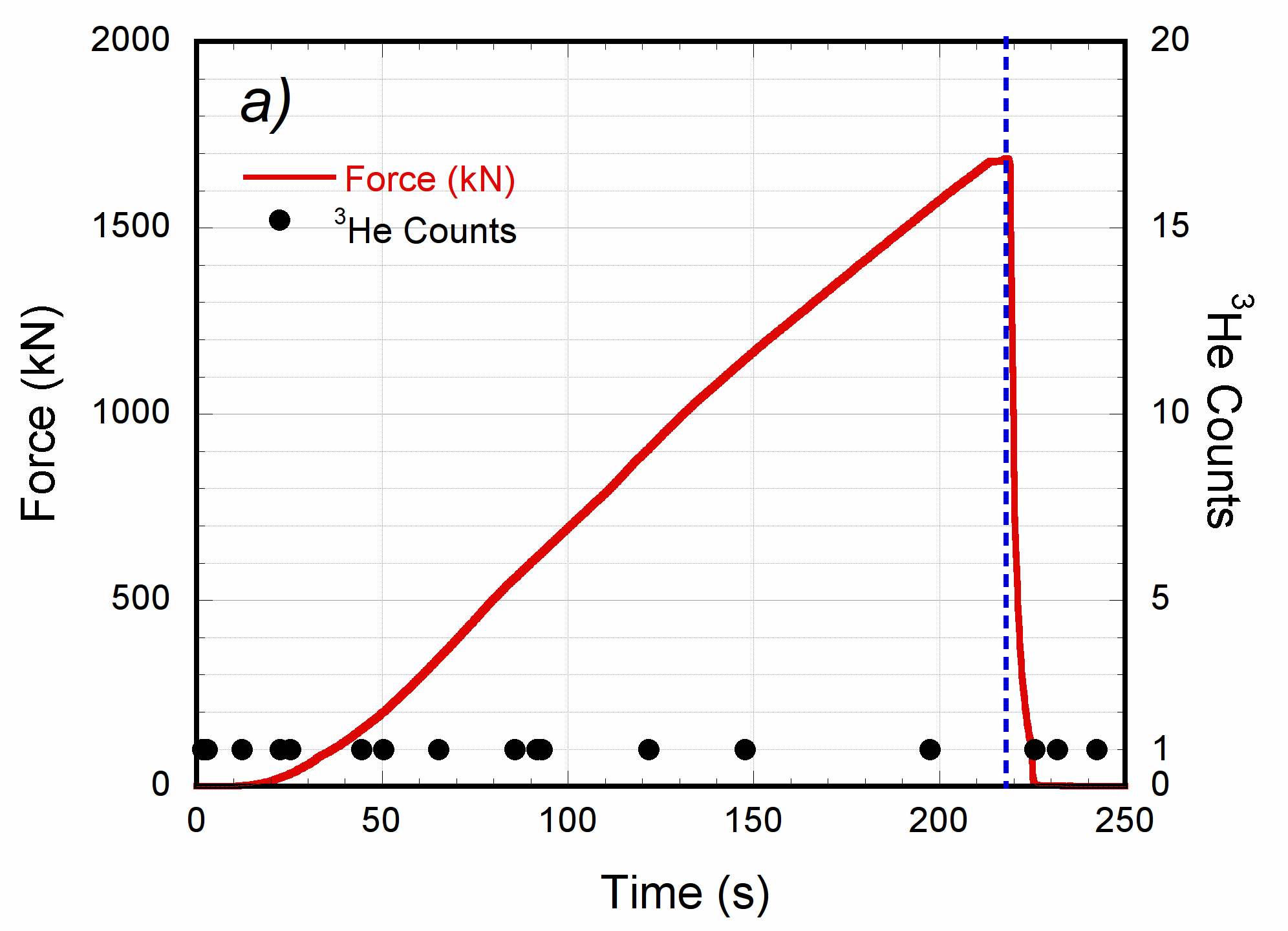}\\
\includegraphics[width=0.85\textwidth]{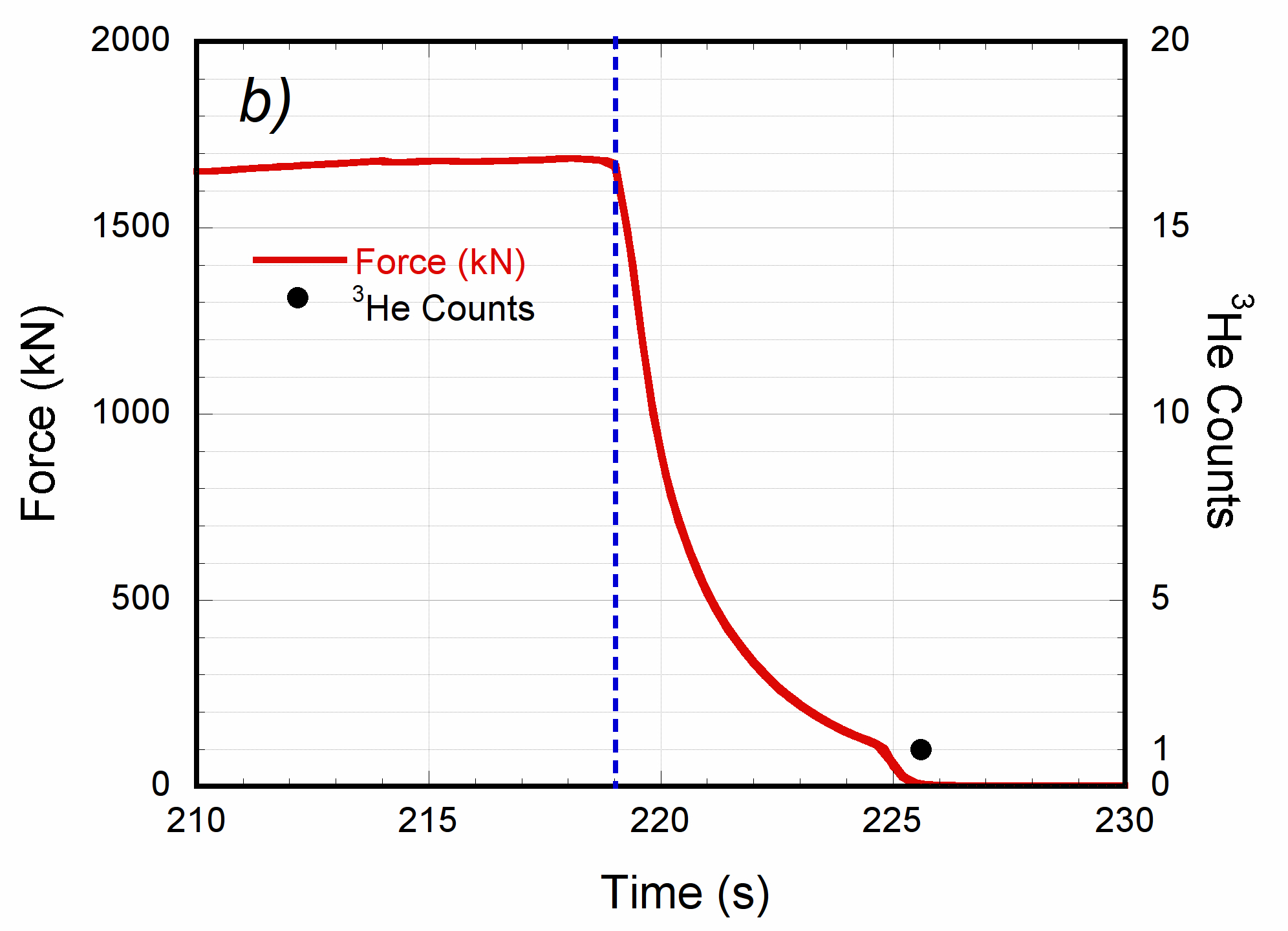}
\caption{$^3$He counts (black dots) together with the loading force on the specimen (red line) for the test with the whole neutron detector protected against acoustic effects. In a) the whole experimental test time is plotted; in b) a zoom around the crashing time is represented. The vertical blue dashed line, in the plots, indicates the crashing time.}
\label{Fig6}
\end{center}
\end{figure}

Eventually a suitable protection was achieved, i.e. no pulses observed above the background ones, and test on granite restarted. This protection, as shown in Fig. \ref{Fig5}, was made of tubular foam covering the $^3$He tube, its cable and the preamplifier. Furthermore, the detector was wrapped with a pluriball foil.
Of course we were aware that the gun shot wasn't necessarily equivalent to that generated in the collapse of the granite specimen, but we considered a success to have at least eliminated this kind of perturbation. As matter of facts, after then we never observed a neutron signal in conjunction with the collapse of granite blocks, except the single pulse background. A record of a typical test is reported in Fig. \ref{Fig6}.

\begin{figure}[p]
\begin{center}
\includegraphics[width=0.85\textwidth]{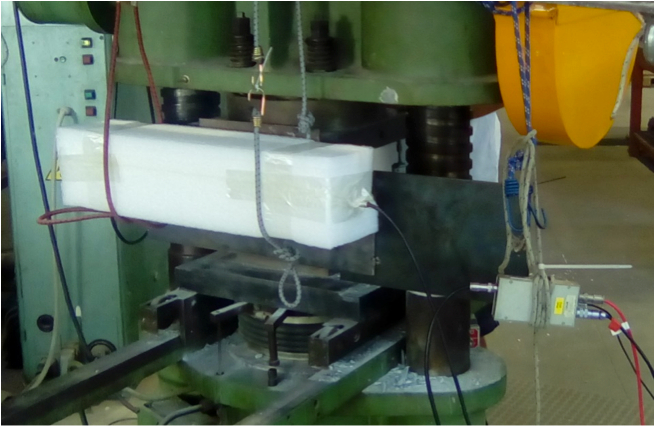}
\caption{The suspended detector, embedded in polystyrene protection and the preamplifier, now unsheltered.}
\label{Fig7}       
\end{center}
\end{figure}

Once ascertained that with a suitable protection of the detector, at the time of the collapsing of the Luserna granite, no signal occurred, neither from detected neutrons nor from acoustical effects, it was decided to repeat tests without acoustic protection, except for the $^3$He tube, where the protection was replaced with a polystyrene shield, as shown in Fig. \ref{Fig7}. 
Apparently this last configuration had also been adopted in the work of Carpinteri et al. \cite{Carpinteri}.

\begin{figure}[h]
\begin{center}
\includegraphics[width=0.85\textwidth]{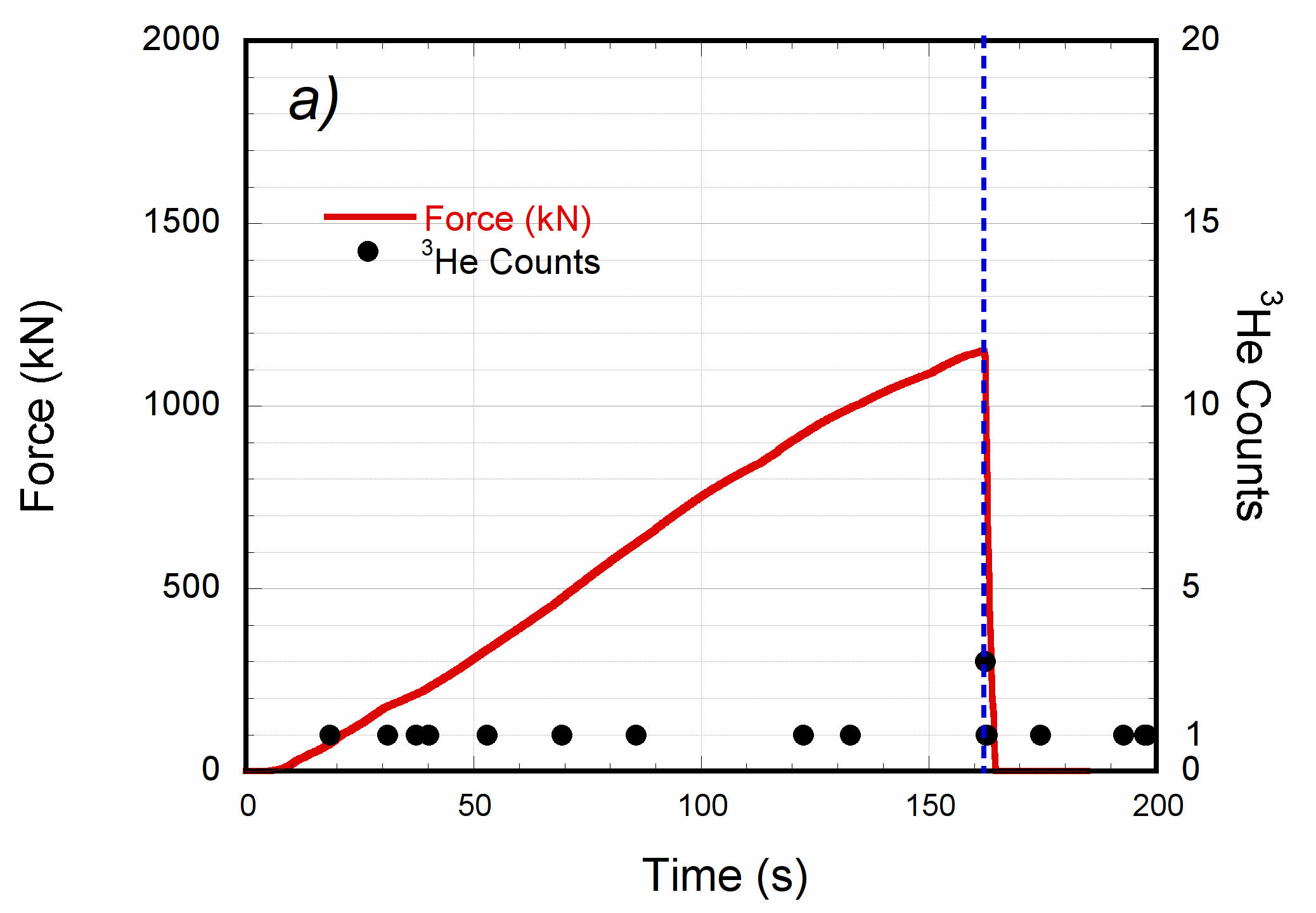}\\
\includegraphics[width=0.85\textwidth]{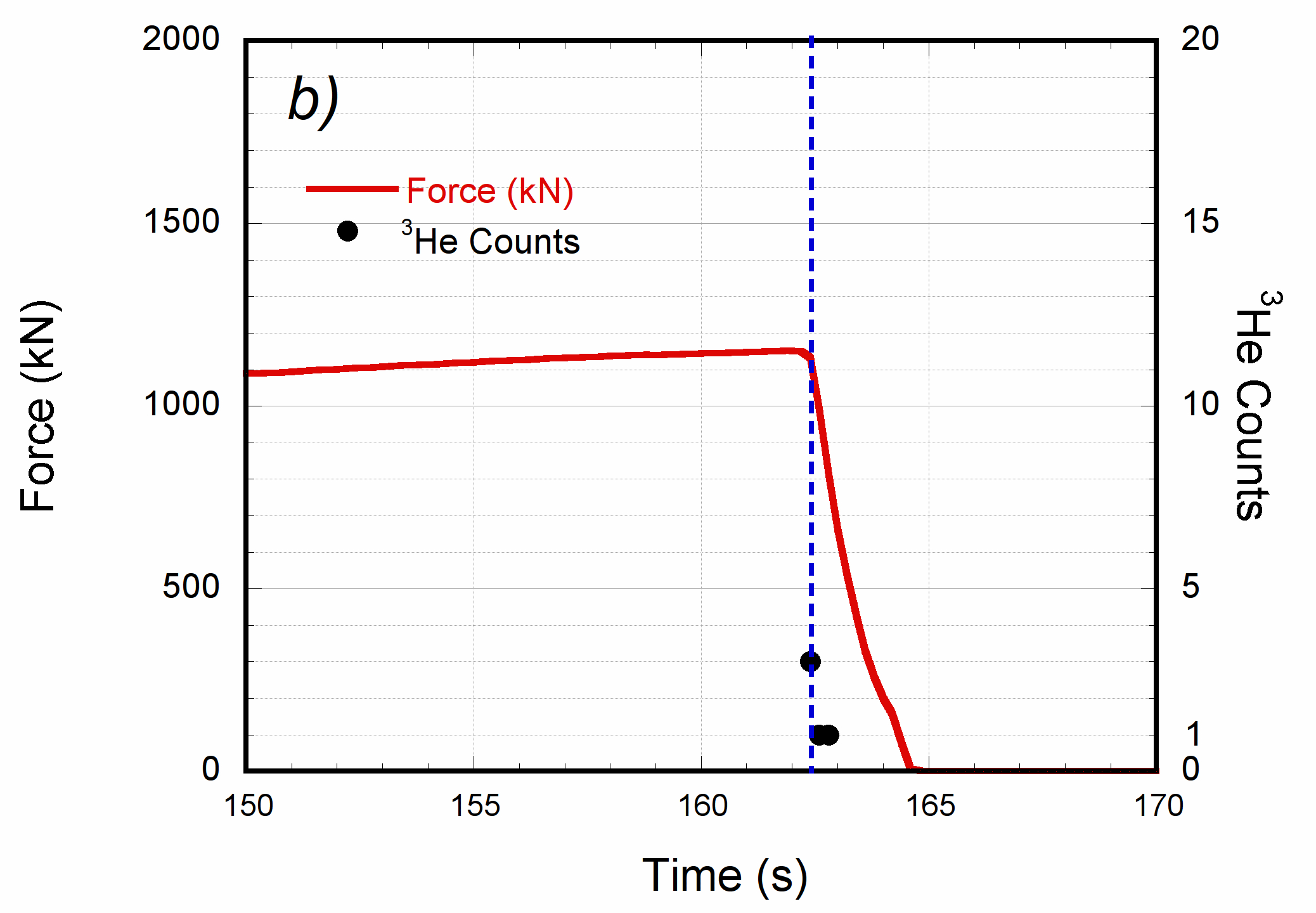}
\caption{$^3$He counts (black dots) together with the loading force on the specimen (red line) when only the $^3$He detector is protected by a polystyrene shield, as by a way in \cite{Carpinteri}. In a) the whole experimental test time is plotted whereas in b) a zoom around the crashing moment is represented. The vertical blue dashed line, in the plots, indicates the crashing time.}
\label{Fig8}
\end{center}
\end{figure}

Not only the system was now sensitive again to the blank gun but also at the crashing time of a granite block the TTL signals appeared again (Fig. \ref{Fig8}) confirming that even the preamplifier/cable are sensitive to the acoustic noise.

\section{Conclusions}
\label{Conclusions}

Tests have been performed to verify the results reported in the paper
of Carpinteri et al. \cite{Carpinteri}, namely the emission of neutrons in the collapsing of blocks of Luserna granite under mechanical loading. No neutrons were detected in our experiments, besides the background ones. 

Instead spurious signals were observed at the crashing time, altogether similar to the neutrons ones, 
but they were rejected because their origin had been demonstrated to be the acoustic noise coming from the 
sudden collapse of the granite, affecting the detector-preamplifier assembly.

\section{Acknowledgements}
%\begin{acknowledgements}
%This research did not receive any specific grant from funding agencies in the public, commercial, or not-for-profit sectors.
The authors of this paper thank Stefano Agosteo and Giovanni D'Angelo of the Energy Department of the Politecnico di Milano for the fruitful technical collaboration and discussion. The ''Laboratorio Prove Materiali e Strutture'' of the Department of Civil Engineering and Architecture of the University of Pavia is acknowledged for having made available the staff and the press equipment for the execution of the experimental tests. 
%\end{acknowledgements}

\section*{References}

\bibliography{bibfile}


\end{document}